\documentstyle[aps]{revtex}
\draft
\tighten

\begin{document}

 \preprint{Version 2.0}

\title{Two-body effects in the decay rate of atomic levels}

\author{S. G. Karshenboim\footnote{Electronic address:
sgk@onti.vniim.spb.su}
}
 \address{
 D. I.  Mendeleev Institute for Metrology,
 198005 St. Petersburg, Russia}

\maketitle

\begin{abstract}
Recoil corrections to the atomic decay rate are considered in the order of
$Z\,m/M$. The expressions
are treated exactly without
any expansion over $Z\,\alpha$.
The expressions obtained are valid
both for muonic atoms (for which they
contribute on the level of a few percent
in high $Z$ ions) and for electronic atoms.
Explicit results for Lyman-$\alpha$ transitions for low-Z
of the order  $(Z\,m/M)\,(Z\,\alpha)^2$ are also presented.
\end{abstract}


\large

This work is devoted to a calculation of the recoil effects for the atomic
decay rate. These effects can be more important in muonic atoms.
The leading term is due to nonrelativistic emission by the nucleus and
it has relative order $Zm_{\mu}/M_N$ or approximately $1/20$ (in light
muonic atoms, for which $A\simeq 2Z$). The nuclear emission was first
taken into consideration in Ref.
\cite{Frid} for hydrogen-like systems. That result is actually a
non-relativistic one, which
includes all corrections in the expansion
over both recoil small parameters
$m/M$ and $Z\,m/M$. It is of the form
\begin{equation} \label{Glead}
\Gamma_{NR-Rec}=\Gamma_\infty\cdot\frac{m_R^3}{m^3}\cdot\left(
1+Z\,\frac{m}{M}\right)^2,
\end{equation}
where $\Gamma_\infty$ is the non-relativistic
decay rate of a level in an atom
with an infinite nuclear mass
within the dipole approximation.
Some development and discussions were
also presented in Refs. \cite{Bac,Dra86}.
The result was deduced there from the classical non-relativistic
two-particle hamiltonian. As it was found in Ref. \cite{JETP95}
it can be also obtained from dimensional analysis and symmetry between
the electron and the nucleus with spin $1/2$.  It was also demonstrated
there that the result for the decay rate is directly connected with
that for the Bethe logarithm ($\log(k_0)$), which is well known 
\cite{comment}.

The approach of Ref. \cite{JETP95} is adjusted here for the relativistic
correction in the expansion over $Zm/M$. The important point of 
consideration
there and here is that the first order $Zm/M$ correction to the dipole 
matrix
element has to have a spin indepen\-dent form. Hence, it is enough to 
evaluate
the expression for a nucleus with spin $I=1/2$.
Some symmetry between the electron
and nucleus has to be in the leading non-relativistic expression.
The general expression is not symmetric in the
relativistic case. In that case,
the nuclear wave function is a free spinor, but the electronic one is an
eigenfunction of the Dirac equation. The expressions to derive
are valid for both the electronic and muonic system,
but in  the last case, they
are expected to be more important for the comparison between theory and
experiment.

The emission of photon in $\lambda' \to \lambda$ transition in 
(electronic)
hydrogen-like atoms is due to a matrix element of the electronic current
\begin{equation}  \label{current}
   J_\mu^{\mbox{(exact)}}(\lambda' \to \lambda) =
\int{d^3 r_e\,
\Psi^+_{e\lambda}({\bf r_e})
\,W^+_N
  \bigg( e\, \gamma_0^e\gamma_\mu^e\, e^{-i{\bf q}\,{\bf r_e}}
-Ze\, \gamma_0^N\gamma_\mu^N \bigg)
\,W_N \, \Psi_{e\lambda'}({\bf r_e})}
,\end{equation}
where $\Psi_e$ and $W_N$ are the electron
and nucleus wave functions,
respectively. The indexes $e$ and $N$ indicate
the particle, $Z$ is the nuclear
charge, and relativistic units in which
$ \hbar  = c = 1$ and $ \alpha=e^2 $ are
used.
The superscript `exact' means that the expression is an exact
relativistic one, but only corrections due to expansion over a small
parameter $Zm/M$ are included. The corrections proportional to $m/M$
are neglected. In that approximation the nucleus is in the origin
(${\bf r}_N=0$), and its wave function $W_N$ is a free
nuclear spinor with using the atom rest frame condition: 
${\bf p}_N=-{\bf p}_e$.
Nuclear multipole moments like the anomalous magnetic dipole
moment or electric quadrupole moment are omitted in the expression above
and if it were necessary they can be taken into consideration 
additionally.

The decay rate to be calculated is indeed a gauge
independent value, but the intermediate results may be gauge dependent
and we are working here in the Coulomb gauge, in which only 
the spatial components
of the current are important. Neglecting here nuclear spin
effects\footnote{Any result is averaged over the nuclear spin.}, 
one can easily
obtain
\begin{equation}  \label{current1}
  {\bf J}^{\mbox{(exact)}}(\lambda' \to \lambda) = e\,
\int{d^3 r\,
\Psi^+_\lambda({\bf r})
  \left( \bbox{\alpha}\,e^{-i{\bf q}\,{\bf r}} +  \,\frac{Z}{M}\,
{\bf p} \right) \,
\Psi_{\lambda'}({\bf r})}
.\end{equation}
The result is directly derived from the free nuclear spinor approximation
within the atom rest frame
condition. Now any value corresponds to the electron
and index `e' may be omitted.

Following Ref. \cite{JETP95} we are reproducing
the leading correction evaluation.
The exponent can
be substituted for unity, and the electron wave function can be 
approximated
by the product of the free spinor and the Schr\"odinger Coulomb wave 
function.
Neglecting electron spin effects, one can find
within the dipole approximation
\begin{equation}         \label{lead}
  {\bf J}^{(0)}(\lambda' \to \lambda) = e\,
\int{d^3 r\,
{\Phi}^*_\lambda({\bf r})
  \left[ \bigg(\frac{1}{m}+ \frac{Z}{M}\bigg)
{\bf p}\right] \,
\Phi_{\lambda'}({\bf r})
}
.\end{equation}
Corrections to the frequency are of the order  $m/M$ and to be
neglected and one can return to Eq. (\ref{Glead}).

Now we evaluate the exact expression.
The corrections to the leading term of Eq. (\ref{lead})
are due to a factor
of $\big(e^{-i{\bf q}\,{\bf r}}-1\big)$ and the difference between exact
Dirac wave function and the approximation used. With using the
anticommutator
\begin{equation}  \label{anti}
  \left\{ {\cal H}_D, \bbox{\alpha} \right\} = 2{\bf p} -
2 V_C \bbox{\alpha}
,\end{equation}
where ${\cal H}_D$ is the Dirac Hamiltonian of the electron in the 
external
Coulomb field
 \begin{equation} {\cal H}_D = \bbox{\alpha}{\bf p} + \beta
m + V_C({\bf r})
 ,\end{equation}
the current of Eq. (\ref{current1}) can be rewritten in the form
 \begin{equation}  \label{curg2}
 {\bf
J}^{\mbox{\rm (exact)}}(\lambda' \to \lambda) = e\, \int{d^3 r\,
  \Psi^+_\lambda({\bf r})\, \bbox{\alpha} \left\{ \left[1 +  
\,\frac{Zm}{M}
\right] + \frac{Zm}{M}\,
\left[ \bigg(e^{-i{\bf q}\,{\bf r}}-1\bigg) + \frac{
\left(\epsilon_{\lambda'} + \epsilon_{\lambda} - 2 V_C \right)}{2m} 
\right]
 \right\} \Psi_{\lambda'}({\bf r})} .
 \end{equation}

We divide above the current into two parts. The first term
includes the leading term, and the other
is only a recoil correction.

We can now start to evaluate the relativistic recoil correction
of relative order $Zm/M(Z\alpha)^2$.
Results will be presented as relative coefficients 
$C_{Rel}$ and $C'_{Rel}$, which are 
implicitly defined by
\begin{equation}
\Gamma=\Gamma_\infty\,\cdot\,\left(1+Z\,\frac{m}{M}\right)^2
\cdot\left\{1+C_{Rel}\,(Z\alpha)^2+C'_{Rel}\,Z\,
\frac{m}{M}\,(Z\alpha)^2\right\},
\end{equation}
which is very convenient for further discussion.
In this definition the recoil term ($C'_{Rel}$) is due
to the second term of Eq. (\ref{curg2}) and its
evaluation is simpler
than in case of
a non-recoil relativistic one ($C_{Rel}$). One can see that
this contribution has just order of  $Zm/M(Z\alpha)^2$ and
that allows to approximate electronic wave function by
the product of free spinor and Schr\"odinger function
\begin{equation}  \label{curg3}
  J_j^{(2)}(\lambda' \to \lambda) = \frac{Ze}{M}\,
\int{d^3 r\,
\Phi^*_\lambda({\bf r}) \,
\frac{1}{2} \Big\{p_j, \,
\left[ \frac{\Big({\bf q}\,{\bf r}\Big)^2}{2}
 + \frac{\left(\epsilon_{\lambda'} + \epsilon_{\lambda} - 2 V_C
 \right)}{2m} \right]
\Big\}
\,\Phi_{\lambda'}({\bf r})}
.\end{equation}

This expression is $j$-independent. The evaluation for the
Lyman-$\alpha$ transition can be simplified acting by operator $p_j$
on $\Phi_{1s}({\bf r})$. The radial integrals appeared after that
\[
I_k=\int_{0}^{\infty}{dz\,e^{-(1+1/n) z}\, z^k\, F\big(2-n,4,2k/n\big)}
\]
are found using Appendix f of Ref. \cite{III}
\[
I_k= k!\,\left(\frac{n}{n+1}\right)^{k+1}
\,{}_2F_1\big(2-n,k+1,4,2/(1+n)\big)
.\]

The explicit results for the
$2p$ states in the hydrogen-like atoms are presented in the Table
\ref{tbl1}, and
for  $np\to 1s$ transitions the general expression is of the form
\begin{equation}
\frac{\Delta \Gamma}{\Gamma} = \frac{Z m}{M}\,(Z\alpha)^2\,
\frac{
\frac{n^2}{10}\left(n^2-1\right)^2\,I_5-4\,
\left(n^2+1\right)\,I_3+n^2\,I_2+2\,n^2\,I_1}{
\left(n^2-1\right)I_4}
.\end{equation}

In conclusion it should be noted that
this work is rather devoted to muonic atoms
because the corrections considered above
and summarized in Table \ref{tbl1}
are
more important there and enter at the level of a few percent
for high nuclear charge $Z$.
The expressions obtained in this work
are of use both for muonic and electronic
hydrogen-like atoms. In the case of ``usual'' (electronic) atoms
some of them are valid also for few electron high-Z systems.
The results are obtained above for the decay rate, but they can be easily
adapted for the line intensity
\[
\frac{\Delta I}{I} = \frac{\Delta \Gamma}{\Gamma}
,\]
because the frequency can include only corrections due to
expansion in $m/M$, but not in $Zm/M$.

A part of this work was supported by Max-Planck-Institut f\"ur Physik
komplexer Systeme and  Institut f\"ur Theoretische Physik der Technische
Universit\"at Dresden
during my stay there. I am very grateful to Profs. G. Soff and U. 
Jentschura
for their warm hospitality and their interest to my work.
V. Ivanov, H. Pilkuhn and A. Yelkhovsky are gratefully aknowledged for
useful discussions. A
part of the work was also supported by the grant \#95-02-03977 of the
Russian Foundation for Basic Research.

\newpage

%
% Table 1
%

\begin{table}[htb]
\begin{center}
\begin{minipage}{7.0cm}
\renewcommand{\arraystretch}{2.0}
\begin{tabular}{ccc}
{states}
& $C_{Rel}$
& $C'_{Rel}$
\\
\hline
\hline
$2p_{1/2}$ & $\ln\frac{9}{8}$ & $1$ \\
\hline
$2p_{3/2}$ & $-\frac{7}{48}-\frac{1}{2}\ln\frac{32}{27}$ & $1$ \\
\end{tabular}
\end{minipage}
\end{center}
\caption{\label{tbl1}  Leading relativistic and recoil corrections for the
$2p \to 1s$ transition.}
\end{table}


\begin{thebibliography}{9}
\bibitem{comment} As it is well known, the Bethe logarithm for a state 
$\lambda$
can be presented as 
a sum over all discrete and continuous intermediate states 
%
\[
\frac{4}{3}\,\frac{\alpha}{\pi}\,\frac{(Z\alpha)^4}{n_\lambda^3}\log{
\big(k_0({\lambda})\big)}=
-\frac{1}{2}\,\sum_{\lambda'\neq\lambda}
{\,\left(
\frac{4}{3}\, \alpha \, \omega_{\lambda'\lambda}^3\,
|{\bf r}_{\lambda'\lambda}|^2\right)\, 
\log{\frac
{2| \omega_{\lambda'\lambda}|}
{(Z\alpha)^2}
}} ,  
\]
%
where the natural designations for the radial matrix element, the 
difference 
of energies and the 
principal quantum number are used. The factor in brackets under the 
summation 
 differs from the decay rate $\Gamma(\lambda\to \lambda')$
only because it is defined for any states, and
the decay rate propertly is defined only for positive 
$\omega_{\lambda'\lambda}$. Due to the
connection between these two values one can reproduce the calculation of 
the Bethe 
logarithm coefficient and will obtain a factor $(1+Zm/M) (m_R/m)^3$ for it
(see Ref, \cite{JETP95} for details).
\end{thebibliography}
\end{document}